\documentclass[prb,showpacs,a4paper]{revtex4}
\usepackage{graphicx} 
\usepackage{graphics}
\usepackage{dcolumn}  
\usepackage{bm}       
\usepackage{amsmath}
\usepackage{amsfonts}
\usepackage{amssymb}
\usepackage{epsfig}
\begin{document}
\pdfoutput=1
\title{Coulomb interaction and charge neutrality: Pariser, Parr and Pople Hamiltonian versus the  Extended Hubbard Hamiltonian}
\author{E. San-Fabi\'an$^{1,2}$, 
J.A. Verg\'es$^{3}$, G. Chiappe$^{2,4}$, E. Louis$^{2,4}$}
\affiliation{
$^1$Departamento de Qu\'{\i}mica F\'{\i}sica,  Universidad
de Alicante, San Vicente del Raspeig, 03690 Alicante, Spain. \\
$^2$Unidad Asociada del CSIC and
Instituto Universitario de Materiales, Universidad
de Alicante, San Vicente del Raspeig, 03690 Alicante, Spain.\\
$^3$Departamento de Teor\'{\i}a de la Materia Condensada,
Instituto de Ciencia de Materiales de Madrid (CSIC), Cantoblanco, 28049 Madrid, Spain.\\
$^4$Departamento de F\'{\i}sica Aplicada,  Universidad
de Alicante, San Vicente del Raspeig, 03690 Alicante, Spain.}
\date{\today}

\begin{abstract}
The Extended Hubbard Hamiltonian used by the Condensed Matter community   is nothing but a simplified version of the Pariser, Parr and Pople Hamiltonian, well established in the Quantum Chemistry community as a powerful tool to describe the electronic structure of $\pi$-conjugated planar Polycyclic Aromatic Hydrocarbons (PAH). We show that whenever the interaction potential is {\it non-local},  unphysical charge inhomogeneities may show up in finite systems, provided that electrons {\it are  not neutralized} by the ion charges. Increasing the system size does not solve the problem when the potential has an infinite range, and for finite range potentials  these charge inhomogeneities become slowly less important  as the   potential range decreases and/or the system size increases. Dimensionality does also play a major role. Examples in bi-dimensional systems, such as planar PAH and graphene, are discussed to some extent. 
\end{abstract}
\maketitle

\section{Introduction}
\label{intro}
Model Hamiltonians are still valuable tools  in Physics and Chemistry  \cite{PP53,Po53,Hu63,BC95,Ma00}. Outstanding examples of microscopic model Hamiltonians  that explicitly include  electron-electron interactions, are those proposed by Pariser-Parr-Pople (PPP) \cite{PP53,Po53} and by Hubbard (Hu) \cite{Hu63}. While the latter restricts interactions to a local term, the former includes, in addition, the Coulomb interaction. The PPP Hamiltonian has been quite successful in describing   a variety of properties of Polycyclic Aromatic Hydrocarbons (PAH)\cite{PP53,Po53,TS86,MT97,MG97,PS01,RA02,MC05,SS07,SR08,LK09,KF11,VC09,VS09,VS10,VS11}, and more recently is also being applied to extended systems\cite{GS11a,GS11b,SS10}. Despite  the important simplifications inherent to both Hamiltonians, they can only be solved exactly in rather small systems, and exceptionally,
for an infinite chain in the case of the Hu Hamiltonian.
Anyhow, both are being greatly useful in going beyond the highly successful DFT approaches. This is a must when interactions are strong, an area in which many of the most interesting  problems in Physics lie. 

The simplest version of the Hubbard Hamiltonian is actually nothing but the PPP excluding Coulomb interactions. In recent years, it  has been generalized to incorporate non-local interactions up to different extents  
\cite{BC95,Ma00,Ma11,PP00,Hi02,ES99,WS11,DW10,HW12,PG12}. For instance, although in most cases only next-nearest neighbors electron-electron interactions have been included, the full unscreened \cite{Ma00} or screened interaction has been incorporated by several authors \cite{PP00,DW10,HW12}. Moreover, a cut-off has been recently introduced assuming the interaction to be negligible beyond a given distance \cite{WS11}. Anyhow, what is mandatory when non-local interactions are included, is to incorporate, as done in the PPP Hamiltonian, the ionic charges that neutralize the electronic charges \cite{BJ07,GK07,LS12,SS11,TO12,LL06}. If this is not done, as for instance in Refs. \cite{WS11,DW10,HW12,PG12,CV07,AE04,AB13,Je08,LS07,GC13}, unphysical charge inhomogeneities  may show up. The problem may be bypassed increasing the system size, provided that the potential range is not infinite. In this paper we compare the two Hamiltonians, emphasizing that the most consistent  way to proceed whenever  non-local interactions are present in the Hamiltonian, {\it is to include the ionic charges that neutralize the electronic charge} \cite{Ma11}. However, this is not a common practice in the Condensed Matter community. 
\begin{table}
\caption{Parameters (in eV) of the HuEx Hamiltonian reported in \cite{WS11} and of the PPP Hamiltonian given in \cite{VS10}. $V_1$, $V_2$ and $V_3$ are the nearest-neighbors, next-nearest-neighbors and third-nearest neighbors hoppings respectively, either as given in \cite{WS11} (taking into account the partially screened frequency dependent Coulomb interaction calculated from first principles) or introducing the value of $U$ in Eq. (7) \cite{VS10}. } 
\begin{center}
\begin{tabular}{ccc}
 Parameter  &  Ref. \onlinecite{WS11} & Ref. \onlinecite{VS10} \\
\hline
$\epsilon_0$ &    -  &  -7.61 \\
 $t_0$ &  -2.8  & -2.34      \\
 $U$   &   9.3   &  8.29    \\
 $V_1$  &   5.5  &   6.44 \\
 $V_2$  &  4.1  &  4.81 \\
 $V_3$  & 3.6  &   4.35  \\
 \hline
\end{tabular}
\end{center}
\label{table:parameters}
\end{table}

\section{Model Hamiltonians}
\label{sec:1}
\subsection{Pariser, Parr and Pople Hamiltonian}
\label{sec:2}
The model Hamiltonian  proposed by  Pariser, Parr and Pople (PPP model) \cite{PP53,Po53} includes  
local on-site and Coulomb interactions.
The Hamiltonian incorporates a single $\pi$ orbital per atom. The PPP Hamiltonian contains
a non-interacting part $\hat H_{0}$ and a term that incorporates the
electron-electron interactions $\hat H_{I-PPP}$:

\begin{equation}
{\hat H}  = {\hat H_{0}} + {\hat H_{I-PPP}} \;.
\label{eq:H}
\end{equation}

\noindent
\noindent Eventually, a core, constant term may be added to account for  the contribution of core electrons to the total energy \cite{VC09,VS09,VS10,VS11}.  The non-interacting term is written as,

\begin{equation}
{\hat H_{0}}  = \epsilon_0 \sum_{i=1,N;\sigma}
{\hat c}^{\dagger}_{i\sigma} {\hat c}_{i\sigma} +
\sum_{<ij> ; \sigma}t_{ij} {\hat c}^{\dagger}_{i\sigma} {\hat c}_{j\sigma}\;,
\label{eq:H_{0}}
\end{equation}

\noindent
where the operator ${\hat c}^{\dagger}_{i\sigma}$ creates an electron
at site $i$ with spin $\sigma$, $\epsilon_0$ is the energy of the orbital, $N$ is the number of  atoms
and $t_{ij}$ is the hopping between nearest neighbor pairs $<ij>$ (kinetic energy).

In cases where the distance $d_{ij}$ between nearest neighbors pairs $<ij>$
significantly varies over the system, the hopping parameter may be scaled. For instance in some PAH or even in defective graphene the C-C distance may differ from its standard value $d_0$ = 1.41 \AA. In such cases one may use a scaling adequate for $\pi$ orbitals \cite{Pa86},
\begin{equation}
t_{ij}=\left(\frac{d_0}{d_{ij}}\right)^3t_0\;.
\end{equation}
\noindent where $t_0$ is a fitting parameter.
The assumption in using  scaling laws is that the interatomic distance will always be around $d_0$, as it actually occurs in most cases.

The interacting part is in turn given by:

\begin{equation}
{\hat H_{I-PPP}}  =
U\sum_{i=1,N}{\hat n}_{i\uparrow} {\hat n}_{i\downarrow}+{\frac{1}{2}}
\sum_{i,j=1,N;i \neq j}V_{ij} ({\hat n}_i-Q_i) ({\hat n}_j-Q_i)\;,
\label{eq:H_{I-PPP}}
\end{equation}

\noindent
The $Q_i$ in the second term of the r.h.s. account for the ionic charges. We allow the ionic charges to depend on site in order to account for the presence of vacancies, impurities, etc. In the present case $Q_i=1$ for all $i$. $U$ is the on-site Coulomb repulsion and $V_{ij}$ is the inter-site Coulomb repulsion, while the density operator is,

\begin{equation}
{\hat n}_{i\sigma}= {\hat c}^{\dagger}_{i\sigma} {\hat c}_{i\sigma}\;,
\label{eq:density}
\end{equation}

\noindent
and the total electron density for site $i$ is:

\begin{equation}
{\hat n}_i= {\hat n}_{i\uparrow} + {\hat n}_{i\downarrow}\;.
\end{equation}

\noindent

In incorporating the Coulomb interaction $V_{ij}$ one may choose the unscreened Coulomb interaction \cite{Ma00}, although it is a common practice to use some interpolating formula. In the case of PAH that proposed by  Ohno  \cite{Oh64} has a wide acceptance,
\begin{equation}
V_{ij}=U\left[1+\left(\frac{U}{e^2/d_{ij}}\right)^2\right]^{-1/2}\;.
\end{equation}
Using this interpolation scheme implies that no additional parameter is introduced and, consequently,  $U$ remains as the single parameter associated to interactions.

Although the PPP model was solved approximately to investigate
 the electronic structure of complex unsaturated molecules, current computation
facilities allow to obtain exact solutions for small PAH.
Recently, we have refined the value of the parameters
entering the model Hamiltonian to get better agreement with experiment and full
{\it ab initio} calculations\cite{VS10,VS11}. In particular,
the effect of $\sigma$-electrons (not included in the model)
is taken into account with the help of DFT calculations.
Actually, fittings to DFT-B3LYP results \cite{VS10} led to parameters within the 
expected range (see Table 1). The values for  $U$ and  $t_0$ are both close to those currently used for graphene 
(see Refs. \onlinecite{VS10,GS11a,GS11b,SS10}
and references  therein). It is not at all surprising  that this parameter set may also be valid for graphene, a system whose electronic configuration is dominated by $\pi$ electrons.

\begin{figure}
\resizebox{0.5\textwidth}{7cm}{\includegraphics{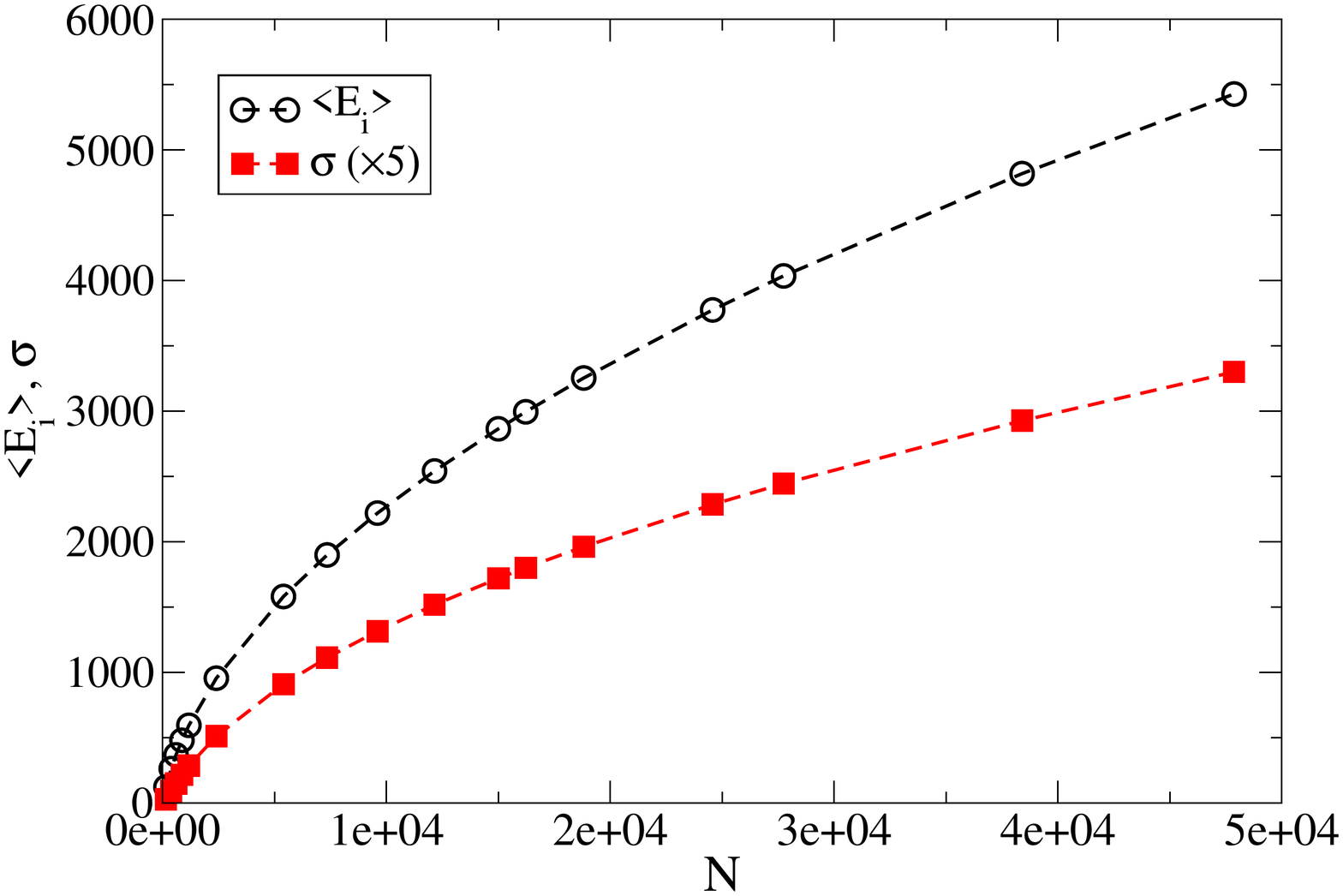}}
\resizebox{0.5\textwidth}{7cm}{\includegraphics{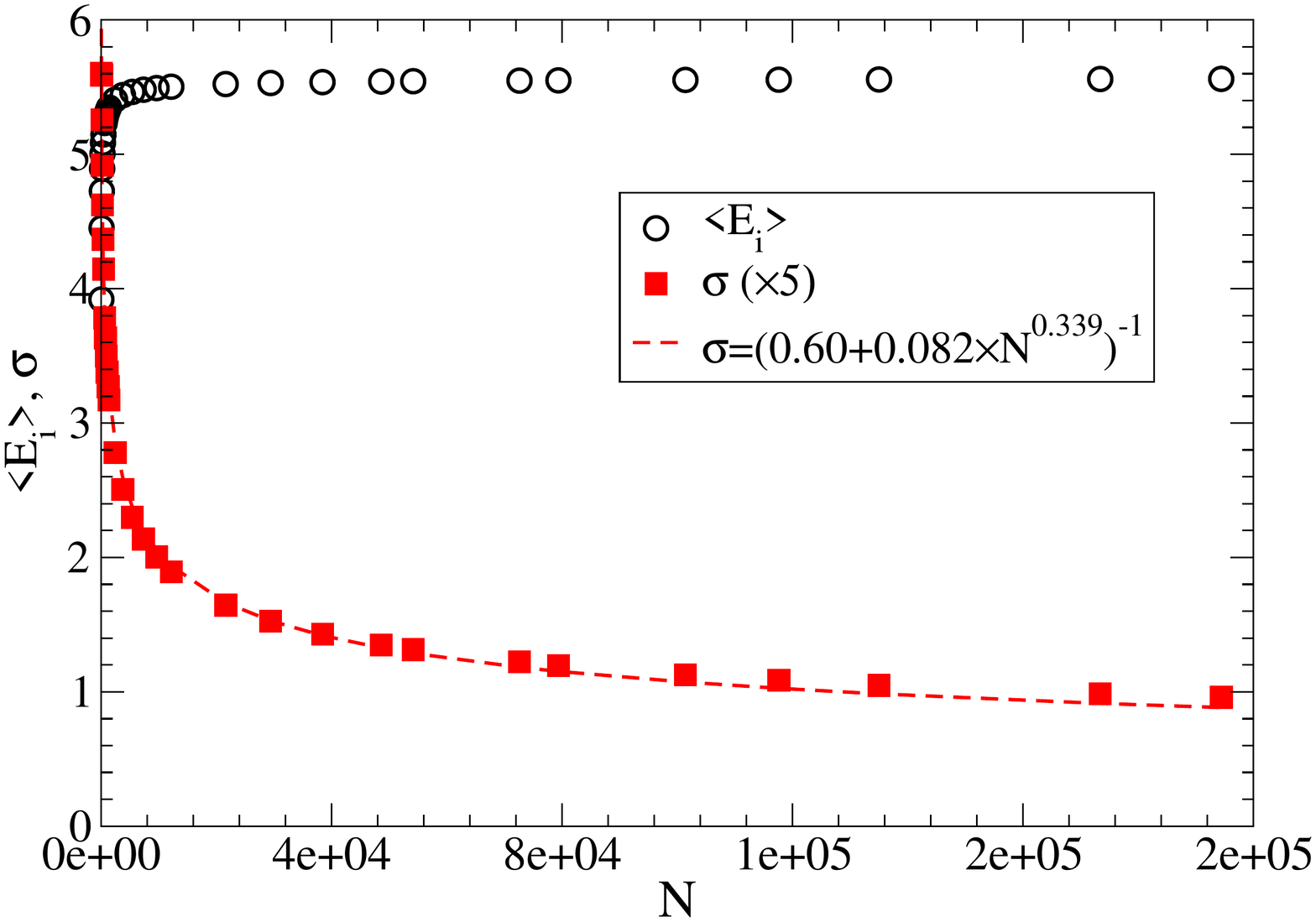}}
\caption{Upper: Average $E_i$ (Eq. (11)) 
 and standard  deviation $\sigma$ of the distribution of $E_i$ (Eq. (10))
calculated by introducing $U$=1 eV  in Eq. (7). Lower: Same as above using the interaction parameters of Ref. \onlinecite{WS11} (see Table 1) divided by $U$. }
\label{fig:1}
\end{figure}

\subsection{Hubbard (Hu) and extended Hubbard (HuEx) Hamiltonians}
The local version of the PPP model ($V_{ij}$=0) is known in Condensed Matter Physics as the Hubbard Hamiltonian \cite{Hu63}. It provides the simplest model describing the effects of electron-electron interaction. On the other hand, the most general version of the extended Hubbard Hamiltonian is commonly written as,

\begin{equation}
{\hat H_{I-Hu}}  =
U\sum_{i=1,N}{\hat n}_{i\uparrow} {\hat n}_{i\downarrow}+{\frac{1}{2}}
\sum_{i,j=1,N; i \neq j}V_{ij} {\hat n}_i{\hat n}_j\;.
\label{eq:H_{I-Hu}}
\end{equation}

Note that  ion charges are not included as.
A considerable confusion exists in the Condensed Matter community because
in some cases\cite{Ma11} the ion charge is included while
in others\cite{Ma00,DW10,WS11,HW12,PG12} it is not. It is likely that the awareness of its major importance is not sufficiently widespread. Remarking on its crucial relevance is the main goal of this paper.

\begin{figure}
\resizebox{0.5\textwidth}{7cm}{\includegraphics{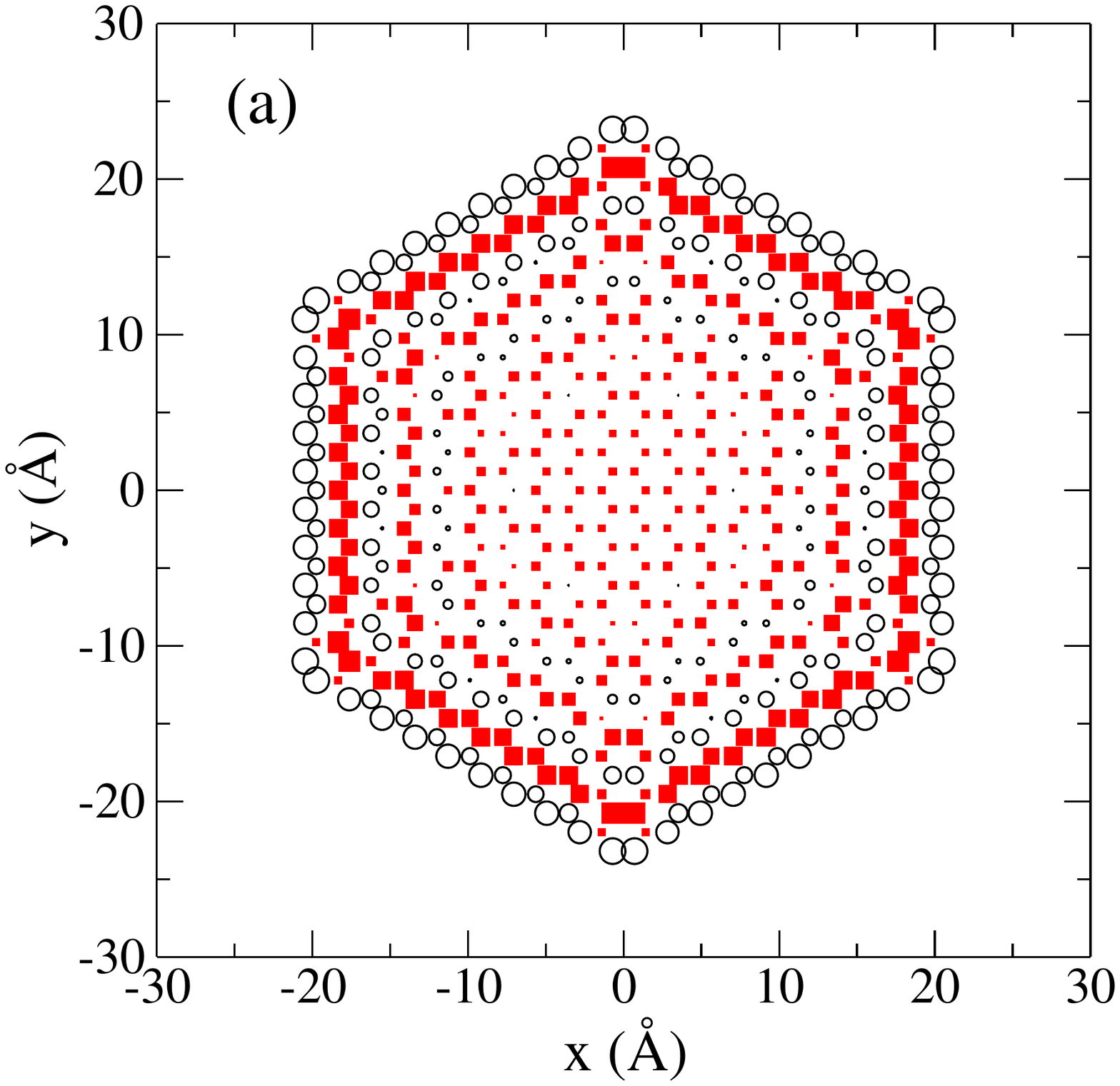}}
\resizebox{0.5\textwidth}{7cm}{\includegraphics{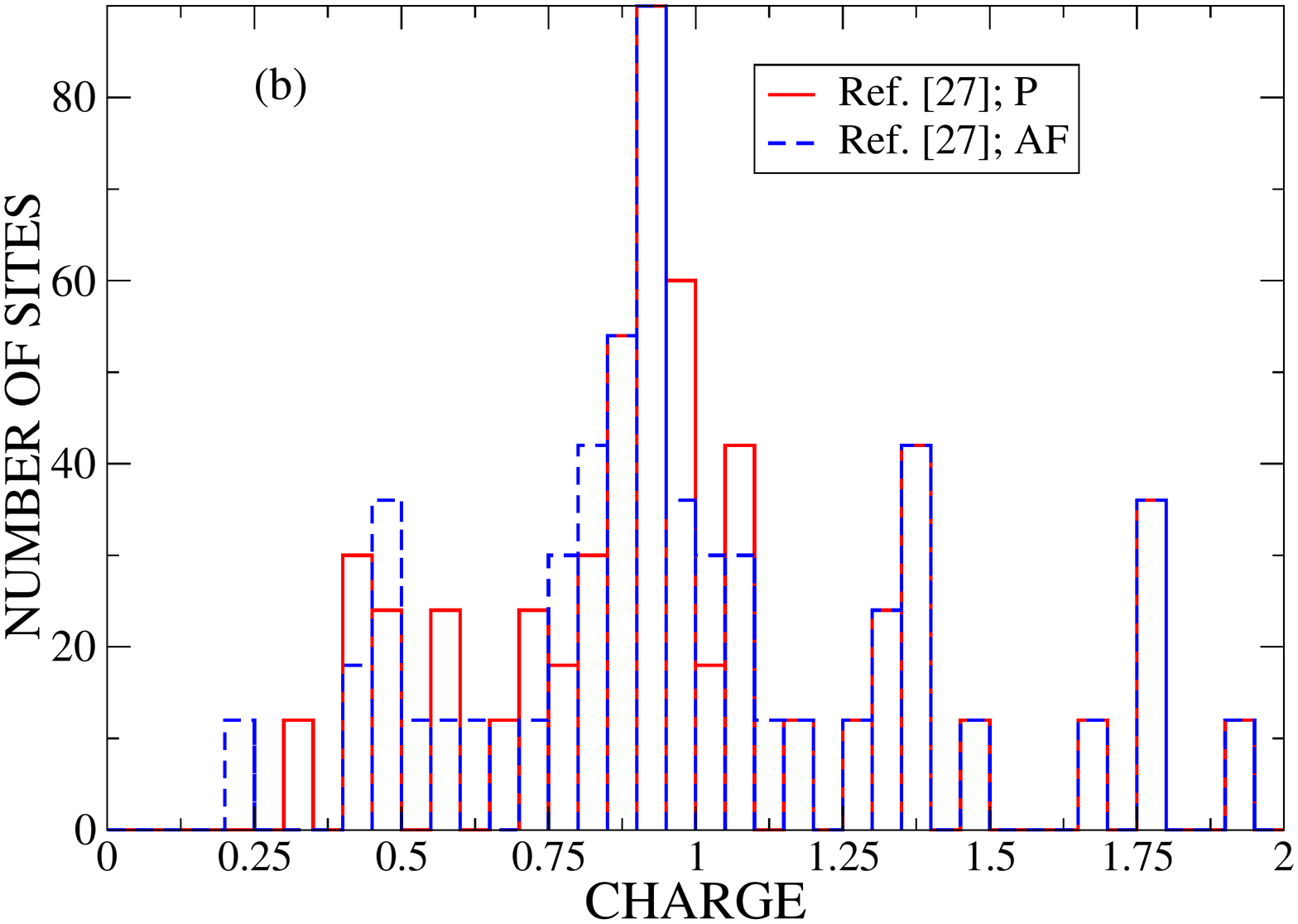}}
\caption{(Color online) (a) $\pi$-electron distribution in a graphene cluster containing 600 C atoms with one electron per atom,  as calculated by means of the extended Hubbard model proposed in Ref. \onlinecite{WS11}. Excess (defect) charges are represented by empty black circles (filled red squares). The results correspond to the paramagnetic configuration.  (b) Histogram of the results depicted in (a) for both the paramagnetic and the anti-ferromagnetic solutions.}
\end{figure}

\subsection{PPP versus HuEx}
To illustrate the effects of the ionic charge we write the difference between the interacting terms of the two Hamiltonians,

\begin{equation}
{\hat H_{I-Hu}} -{\hat H_{I-PPP}} =-{\frac{N}{2}} <E_i> + \sum_{i=1,N}E_i{\hat n}_i\;,
\end{equation}

\noindent
where an effective energy level on site $i$ has been defined by
 
\begin{equation}
E_i = \sum_{j=1,N;j \neq i}V_{ij}\;,
\end{equation}

\noindent
which mean value is
 
\begin{equation}
<E_i >=\frac{1}{N} \sum_{i=1,N}E_{i}\;.
\end{equation}

While the first correction in Eq. (9) is obviously a constant and has no physical consequences,
the second one is site dependent and can only be merged with the orbital energy or
the chemical potential provided that translational invariance is taken for granted.
We show below that due to the fact that the potential is long-ranged,
a translational invariant system is never reached by continuously increasing its size.

\section{Results}
\subsection{Differences between the two Hamiltonians}
In order to illustrate  the effects that the differences between the two Hamiltonians may have, we have calculated $<E_i>$ (see Eq. (11)) and the standard deviation $\sigma$ of the distribution of values $E_i$ given in Eqs. (10),  as a function of the number of sites $N$ in   hexagon clusters of the honeycomb lattice. 
Results obtained by introducing  $U=1$ eV  in Eq. (7) are depicted in Fig. 1.
As expected,  the average    diverges for $N \rightarrow \infty$. The standard  deviation $\sigma$ of the distribution of $E_i$  behaves similarly. This poses serious problems when finite systems are considered, particularly if one is interested in calculating the value of a given magnitude when the system size tends to infinity by means of some finite size scaling procedure. In order to study the case of a non-local but finite range interaction potential we have done a similar calculation for the parameter set of \cite{WS11}, which are actually not that different from those fitted for the PPP Hamiltonian in \cite{VS10} (see Table 1). The results  are also reported in Fig. 1.  Now, as the weight of the surface decreases  the average   $<E_i>$ tends to its bulk value, 5.581, while the standard deviation slowly reaches zero. For system sizes that can be handled in average modern computers (not higher than 10000 sites)  $\sigma$ is still far from zero. 

\begin{figure}
\resizebox{0.5\textwidth}{7cm}{\includegraphics{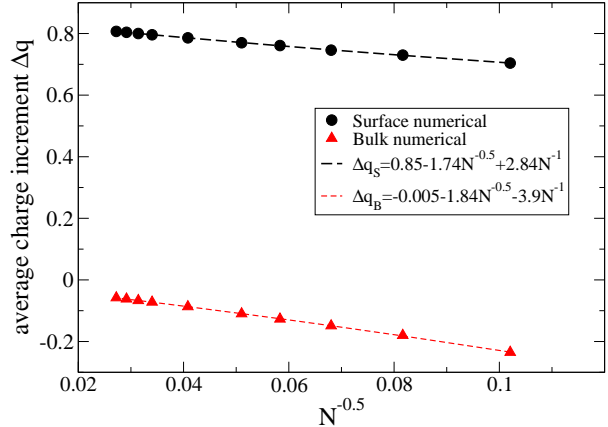}}
\caption{(Color online) Average charge increment (respect to the neutral value of one electron per site) at the surface and $\Delta q_S$ and the bulk $\Delta q_B$ of hexagonal graphene clusters containing  $N$ atoms. Calculations were carried out by means of the  Extended Hubbard Hamiltonians with the parameters of Table I \cite{WS11} whithout  including  the ion charges.}
\label{coronene}
\end{figure}

\subsection{Large Graphene clusters: UHF results}
To illustrate the unphysical consequences of excluding the ion charges we herewith present results for hexagonal clusters of graphene. We use either the PPP Hamiltonian with the parameters of Table 1 or the Hamiltonian proposed in Ref. \onlinecite{WS11} with the model parameters derived taking into account the partially screened frequency dependent Coulomb interaction
calculated from first principles (according to the notation of Ref. \onlinecite{WS11}, parameters cRPA). Note that the interaction parameters $V_1$, $V_2$ and $V_3$ are not very different in the two models.
Both Hamiltonians were solved within the Unrestricted Hartree-Fock approximation. Within UHF, the interacting term of the PPP Hamiltonian is approximated by,
$$
{\hat H}^{UHF}_{I-PPP}=U\sum_{i=1,N;\sigma}\left({\hat n}_{i\sigma} <{\hat n}_{i{\overline \sigma}}>
+\frac{1}{2}<{\hat n}_{i\sigma}><{\hat n}_{i{\overline \sigma}}>\right) $$ 
$$\hspace{-0.5cm} -\frac{1}{2}\sum_{i\neq j}V_{ij}\left(< {\hat n}_i>< {\hat n}_j>-\sum_{\sigma}<{\hat c}_{i\sigma}^{\dagger}{\hat c}_{j\sigma}><{\hat c}_{i\sigma}^{\dagger}{\hat c}_{j\sigma}>-1\right) $$
\begin{equation} -\sum_{i\neq j}V_{ij}\left({\hat n}_i< {\hat n}_j>-\sum_{\sigma}{\hat c}_{i\sigma}^{\dagger}{\hat c}_{j\sigma}<{\hat c}_{i\sigma}^{\dagger}{\hat c}_{j\sigma}>-{\hat n}_i\right)\;. 
\label{eq:H_{I-PPP-UHF}}
\end{equation}
\noindent
A similar equation is valid for the extended Hubbard Hamiltonian with a
small difference consisting in the absence of the third term of both the second and the third parenthesis.

Results are reported in Figs. 2, 3 and 4. It is first noted that excluding ion charges leads to completely unphysical charge fluctuations (up to around $\pm0.9$, see Fig. 2b). Including the ion charges in the same Hamiltonian fully removes charge transfer both in the paramagnetic (P) and in the antiferromagnetic (AF) solutions. We have investigated how the average charge increment (with respect to the locally neutral case of one electron per site) at the bulk and at the surface vary with the number of atoms in the hexagon $N$. The results  for the AF configuration shown in Fig. 3 (very similar results are obtained for the P solution)  indicate that we are not far from what one should expect to be the case for an infinite cluster (the largest cluster in Fig. 3 has 1350 atoms). In the continuum limit, the number of surface atoms can be approximated by $N_S=6L$, where $L$ is the hexagon side  $L\propto N^{-0.5}$, while that of bulk atoms is $N_B=(6\sqrt{3}L)/4$. Charge neutrality implies $N_B\Delta q_B$=-$N_S\Delta q_S$. As the constant term in the fitted curve for $\Delta q_B$ should vanish (actually it is already very low), for an infinite cluster   $\Delta q_B\approx -1.84N^{-0.5}$ and $\Delta q_S\approx 0.85N^{-0.5}$  that can be checked to nearly fulfill charge neutrality.

Local $S_z$ are also significantly changed when the ion charges are included (compare Figs. 4a and 4b) becoming much more similar to the solution obtained with the PPP Hamiltonian (compare Figs. 4b and 4c). These results are for sure a consequence of what was discussed in the preceding subsection and are not at all changed as the size of the system is increased (checks on clusters containing up to around 2000 atoms were carried out).
On the other hand, actual values of model parameters have no effect on the odd results illustrated in Fig. 2a.
\begin{figure}
\resizebox{0.5\textwidth}{7cm}{\includegraphics{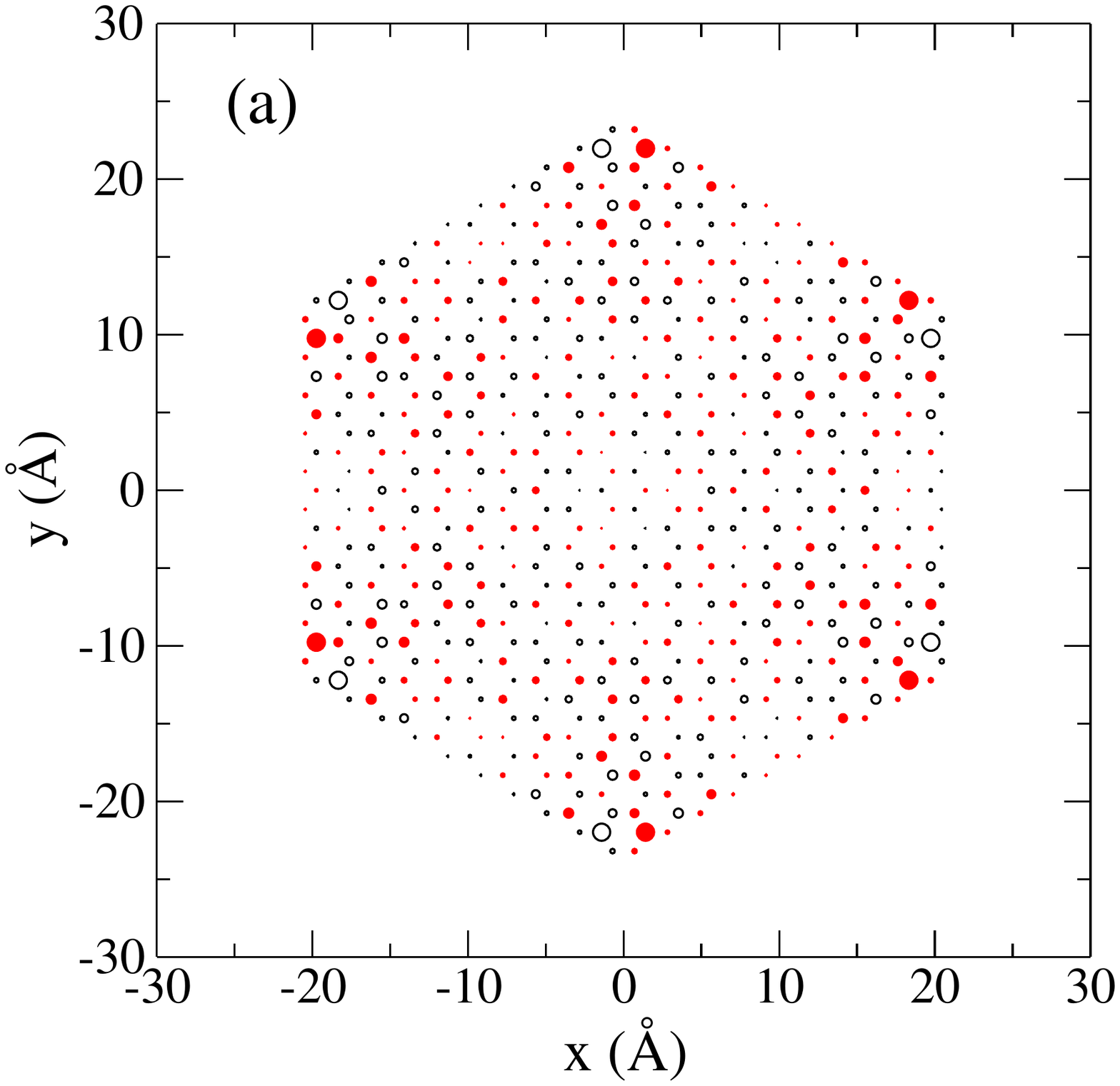}}
\resizebox{0.5\textwidth}{7cm}{\includegraphics{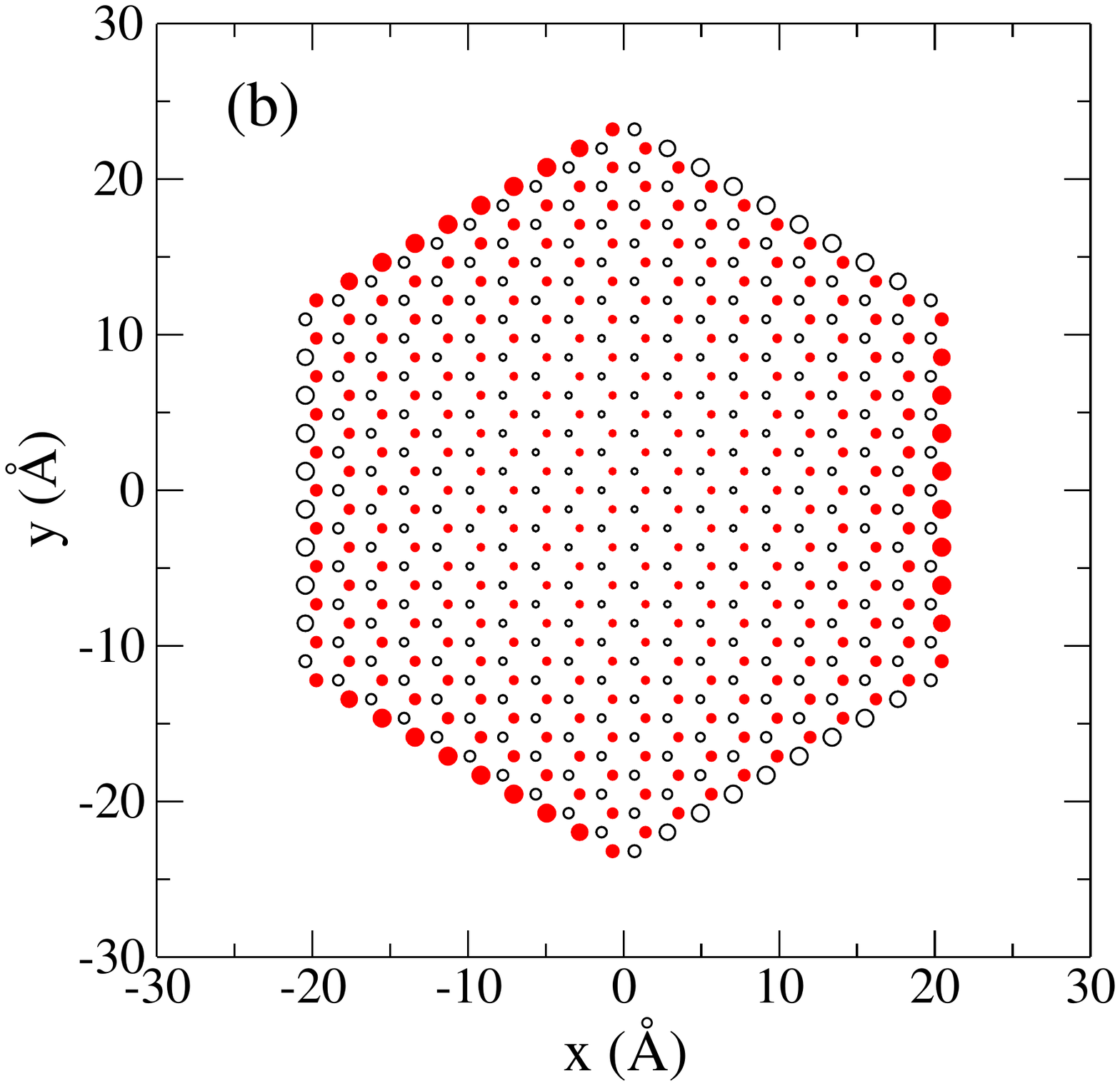}}
\resizebox{0.5\textwidth}{7cm}{\includegraphics{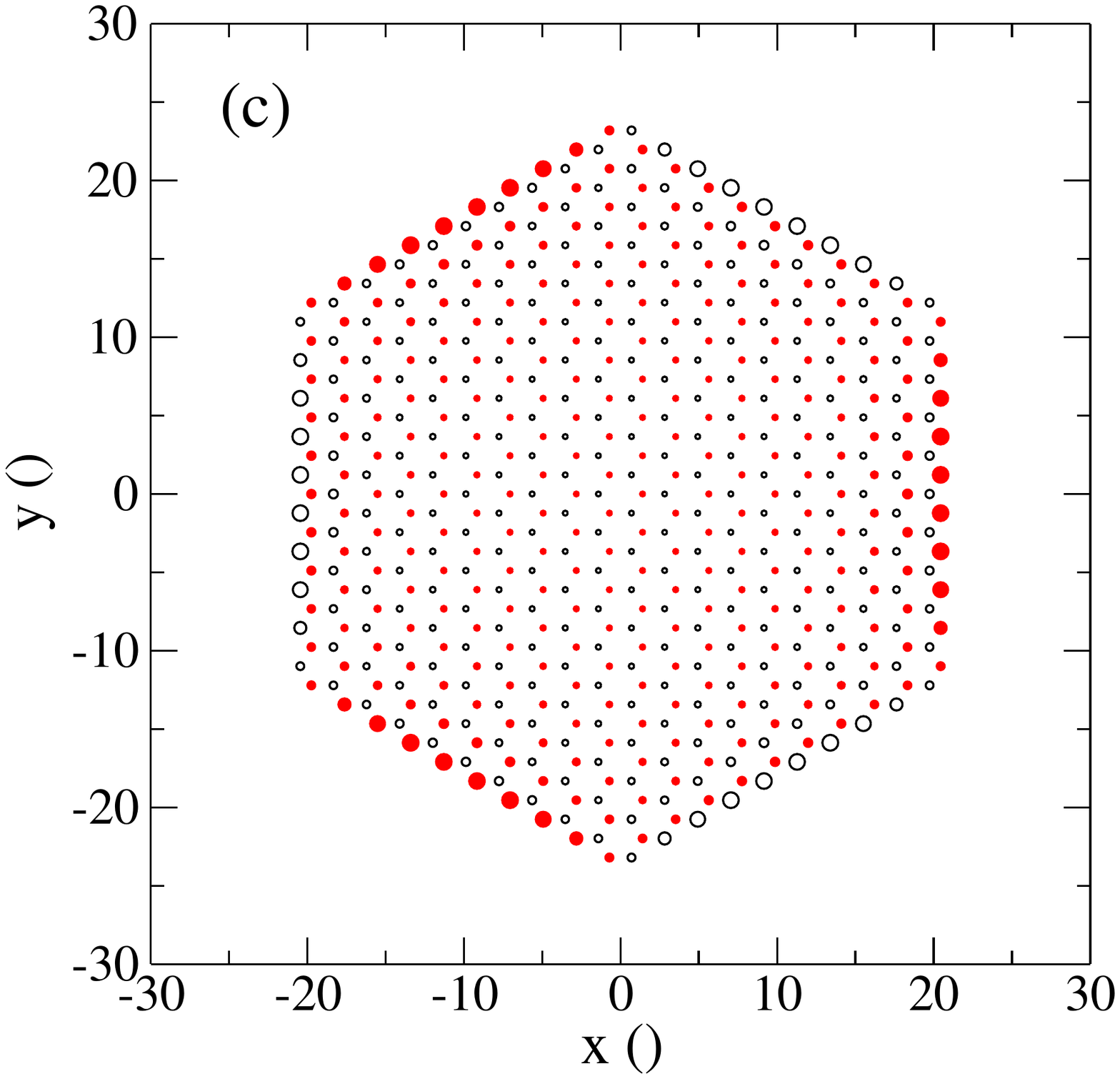}}
\caption{(Color online) (a) Local distribution of the $z$-component of the spin $S_z$ in graphene cluster containing 600 C atoms with one electron per atom. $S_z>0$ ($<0$) are represented by empty black (filled red) circles. Results obtained with the truncated extended Hubbard model proposed in Ref. \onlinecite{WS11}. (b) As in (a) but including the ion charges. (c) As in (a) but calculated with the PPP Hamiltonian with the parameters mentioned in the text.}
\end{figure}

\subsection{Small PAH molecules: {\it ab initio} results}

In this subsection we present results for small PAH molecules calculated by means of the PPP Hamiltonian and of the HuEx Hamiltonian proposed in Ref. \onlinecite{WS11} (see Table 1). In particular we calculate the triplet vertical excitation in anthracene and coronene,
and the  single and double Ionization Energies (IE$_i$, i=1,2) in coronene. In both cases there are   experimental data   available \cite{HS09,AD09,HJ11}.

In solving the PPP and Hubbard Hamiltonians for those two molecules  we use a straightforward Lanczos transformation which, starting from a random ground state candidate, generates a small Hamiltonian
matrix that can be diagonalized to get a better approximation for the
ground state. This process is iterated until convergence is reached (see  \cite{CL93} for details). Actually, as coronene is too large to be solved exactly, we used a recently developed Multi-Configurational (MC) method based upon the just mentioned Lanczos method \cite{VS10}. In some cases, UHF results were also obtained.

As already shown in the previous subsection,  an immediate consequence of excluding ion charges is that the ground-states of small PAH molecules show artificial and large  charge fluctuations. For example,
Fig. 6 shows the groundstate distribution of $\pi$ electrons on anthracene obtained using
the Ref. \onlinecite{WS11} model. If the ion charges are incorporated,   exactly one electron per site is obtained as in the case of graphene discussed above.

\begin{table}
\caption{Triplet vertical excitation energy (eV) of two small PAH obtained for the interaction
models discussed in this work. Data taken from:  [a] Values given in Table XVI of Ref. \onlinecite{HS09}; [b] Values compiled in  Ref. \onlinecite{AD09}; [c] B3LYP parameters of Table III of Ref. \onlinecite{VS10}.}
\label{tab:1}
\begin{tabular}{ccc}
\hline\hline
                         &   Anthracene               &  Coronene                 \\
\hline
experiment               & 1.85-1.87 [a]  & 2.37-2.40 [b] \\
bare                     &   0.02                     &   0.36                    \\
cRPA                     &   0.11                     &   0.37                    \\    
modified bare            &   1.40                     &   2.48                    \\
modified cRPA            &   2.06                     &   2.94                    \\
PPP [c]      &   1.95                     &   2.62                    \\
PPP-UHF [c]  &   1.83                     &   3.02                 \\
Hubbard [c]  &   1.78                     &   2.40                    \\
\hline\hline
\end{tabular}
\end{table}

\begin{figure}
\resizebox{0.5\textwidth}{7cm}{\includegraphics{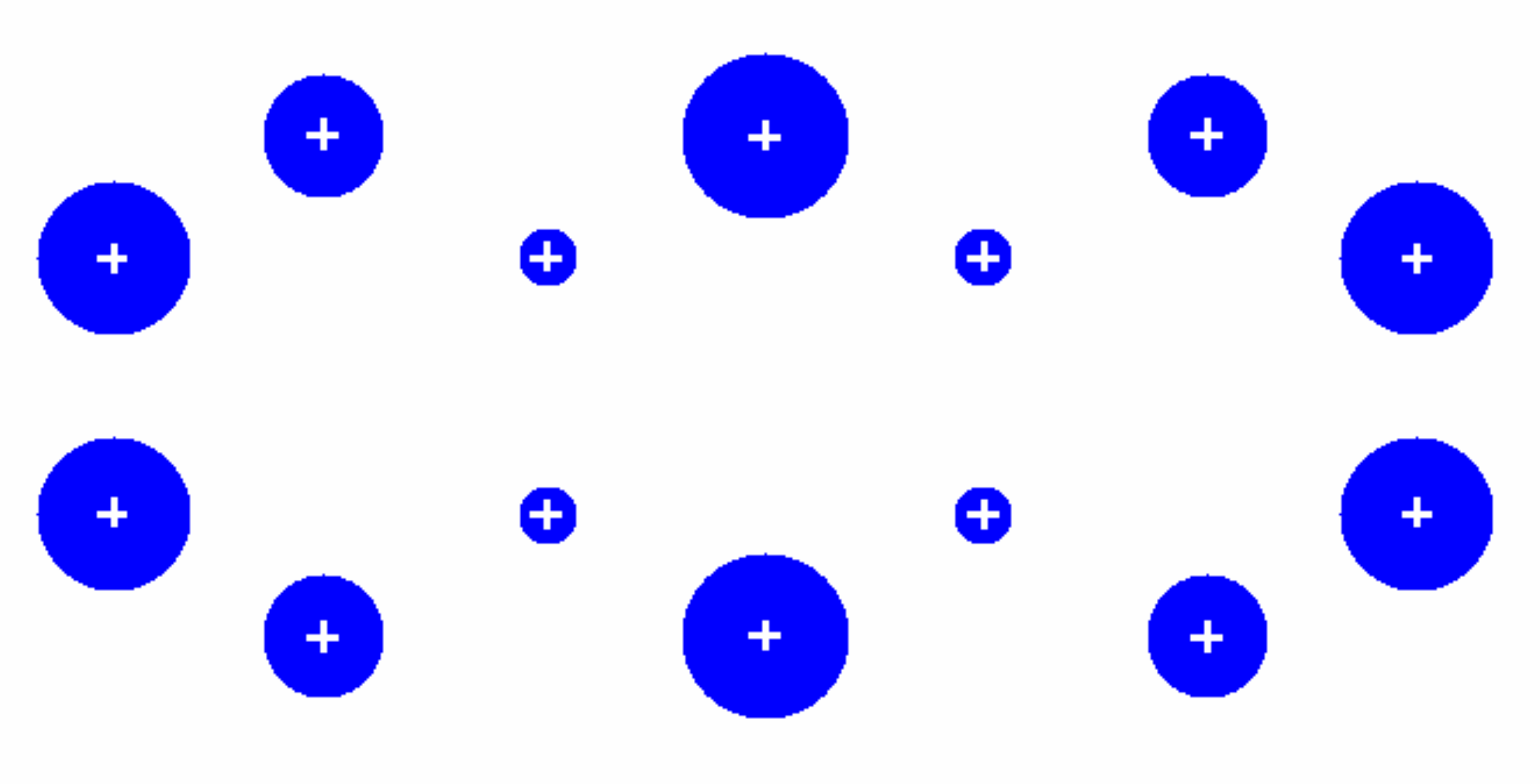}}
\caption{(Color online) $\pi$-electron distribution of neutral anthracene as calculated by
the  Hubbard Hamiltonian of Ref. \onlinecite{WS11}. Nuclei positions are denoted by pluses
while circles areas are proportional to electronic charge.}
\end{figure}

Note that if the ion charges are included into the extended Hubbard model obtained in Ref. \onlinecite{WS11}
it becomes quite similar to PPP model except for the numerical value of the parameters and the
somewhat arbitrary truncation. Notice that Coulomb interaction parameters of PPP model
are usually determined by the value $U$ of the on-site interaction and a well-behaved
interpolation law for the rest that extends interaction to long distances (details can be found in Ref. \onlinecite{VS10}, for
example). This makes sense as long as screening is not metallic.
The question therefore arises of knowing the real predictive value of the extended
model of Ref. \onlinecite{WS11}. Fortunately, we will show now that it works well at least for some fundamental
magnitudes of small PAH molecules. Let us focus to the triplet vertical excitation
energy of anthracene and coronene. It is just the energy difference between the lowest
triplet state and the groundstate, and consequently, easy to calculate and experimentally
well determined. Table I shows that both dressed generalized-extended Hubbard model ({\em after
including ion charges}) and PPP model give reasonable values (ionic relaxation is still missing)
and also to some extent the simple Hubbard model. Non-corrected versions fail completely
to predict this magnitude. To illustrate the rather good performance of UHF, results for PPP-UHF are also shown.

 The extended Hubbard Hamiltonian without ion charges also fails in giving correct ionization energies (IE). Calculating $IE$ requires taking a specific value for the orbital energy. As this parameter was not given in  Ref. \onlinecite{WS11}, we take that used in the PPP Hamiltonian (see above). To illustrate this issue, UHF results suffice. 
Results for coronene, a molecule for which experimental data for single and double Ionization Energies (IE$_i$, i=1,2) are available \cite{HJ11}, are reported in Table II. It is readily noted that while  PPP gives results in agreement both with experiments and DFT calculations, a similar solution of the Hamiltonian proposed in Ref. \onlinecite{WS11} completely fails. Although including ion charges in the
latter Hamiltonian dramatically improves the results, they are still not so good
as those obtained with the PPP Hamiltonian.

\section{Concluding Remarks}

We have shown that the extension of the Hubbard Hamiltonian that incorporates non-local electron-electron interactions may give unphysical results if the ion charges are not included. Although we have concentrated on either small planar PAH molecules or bidimensional graphene clusters, it can be anticipated that  problems should be even more serious  in three dimensional systems. We have shown that the HuEx without  the ion charges gives unphysical charge inhomogeneities both in large graphene clusters and in small PAH molecules. In addition, and for similar reasons, it fails in giving the singlet-triplet excitation energy and the ionization energies of coronene (a molecule that can be considered as one of the smallest graphene clusters). We have also shown that including ion charges dramatically improves all results.  As incorporating the ion charges does not increase the difficulty of the Hamiltonian, we see no reason for continuing using a Hamiltonian that in many cases leads to odd results.

Financial support by the spanish "Ministerio de Ciencia e Innovaci\'on  MICINN" (grants CTQ2007-65218, CSD2007-00006, FIS2008-06743, FIS2009-10325 and FIS2009-08744)
and the Universidad de Alicante is gratefully acknowledged.
We also acknowledge support from the DGUI
of the Comunidad de Madrid under the R\&D Program of activities
MODELICO-CM/S2009ESP-1691.

\begin{table}
\caption{Ionization energies in coronene calculated by means of the UHF approximation  for PPP and HuEx (parameters cRPA) Hamiltonians. Experimental results and DFT results reported in Ref. \cite{HJ11}  are also shown.}
\label{tab:2}
\begin{tabular}{c|ccc}
\hline
   & IE$_1$  &  IE$_1$+IE$_2$  & IE$_3$  \\
\hline
 UHF-PPP &     6.80  & 17.25  & 13.84   \\
 UHF-HuEx & 25.4  & 49.93  &  22.28  \\
 UHF-HuEx (with ion charges) & 5.63  & 12.70  &  8.61  \\
  experimental \cite{HJ11} & 7.29  &  18.7  &  - \\
  DFT \cite{HJ11}   &  7.0 & 17.81  & 14.76   \\
 \hline
\end{tabular}
\label{IE}
\end{table}

%
%

\end{document}